\documentclass[preprint,a4paper,nofootinbib]{revtex4}

\usepackage{xcolor}
\usepackage{amssymb,amsmath}
\usepackage{physics}
\usepackage{graphicx}
\usepackage{hyperref}
\usepackage{derivative}
\usepackage{csquotes}

\newcommand{\gexp}[1]{\overset{\scriptscriptstyle {#1}}{g}{}}

\newcommand{\OrdExp}[2]{\overset{\scriptscriptstyle {#1}}{#2}{}}
\newcommand{\lp}{\left(}
\newcommand{\rp}{\right)}

\begin{document}

\title{Dark Coriolis Fields}

\author{Gabriele Bianchi}
\email{gbianchi1@uninsubria.it}
\affiliation{DiSAT, Universit\`a degli Studi dell'Insubria, via Valleggio 11, Como, Italy \&\\ INFN, Sezione di Milano, via Celoria 16, 20133, Milano, Italy}

\author{Federico Re}
\affiliation{Dipartimento di Fisica \lq\lq Giuseppe Occhialini\rq\rq, Universit\`{a} di Milano Bicocca,\\ Piazza dell'Ateneo Nuovo 1, 20126, Milano, Italy}

\author{Oliver F. Piattella}
\affiliation{DiSAT, Universit\`a degli Studi dell'Insubria, via Valleggio 11, Como, Italy \&\\ INFN, Sezione di Milano, via Celoria 16, 20133, Milano, Italy \&\\ 
Núcleo Cosmo-Ufes, Universidade Federal do Espírito Santo, Vitória, ES, Brazil \&\\
Como Lake Centre for AstroPhysics (CLAP), Universit\`a degli Studi dell'Insubria, via Valleggio 11, Como, Italy}

\date{\today}

\begin{abstract}
     We argue that the standard post-Newtonian expansion scheme used in General Relativity leaves room for time-space components $g_{ti}$ of the metric to be of the same order of the usual gravitational potential. We explore this possibility and find that such leading order contributions to $g_{ti}$ are related to the Coriolis field of Newton-Cartan gravity. We investigate the possibility that Coriolis fields mimic dark matter effects in disk galaxies. We find solutions from their field equations that sustain the velocity rotation curves in the bulge region and beyond it, notably describing flattish velocity profiles. We dub such solutions Dark Coriolis Fields.
\end{abstract}


\maketitle

\section{Introduction}

A generic space-time metric can be written as:
\begin{align}
\label{line element}
    ds^2 &= g_{00}(dx^0)^2 + 2g_{0i}dx^0 dx^i + g_{ij}dx^i dx^j = \nonumber\\
    &= g_{tt}dt^2 + 2g_{ti}dt dx^i + g_{ij}dx^i dx^j\,,
\end{align}%
where $dx^0 = c dt$, so that $g_{tt} = c^2 g_{00}$ and $g_{ti} = c g_{0i}$, and the Latin spatial indices $i,j$ vary from $1$ to $3$. We adopt the convention $(-, +, +, +)$ for the metric, but do not set $c = 1$, as we will be interested in the limit $1/c \to 0$. Since $ds^2$ has the dimensions of a squared length, then $g_{00}$, $g_{0i}$ and $g_{ij}$ are dimensionless, $g_{ti}$ has the dimensions of a speed and $g_{tt}$ of a squared speed. Due to the time-inversion symmetry, the dimensionless $g_{0i}$ must be expanded in only odd powers of $v/c$, while $g_{00}$ and $g_{ij}$ in only even powers. Therefore, $g_{ti} = c g_{0i}$ can contain solely even powers of $1/c$, and the same for $g_{tt}$ and $g_{ij}$.

The conventional post-Newtonian expansion is the following (see, e.g., \cite{Weinberg:1972kfs, Will_2014}):
\begin{equation}
\label{conv post newt}
    g_{tt} = \eta_{tt} + \gexp{0}_{tt} + \gexp{2}_{tt} + \dots\,, \qquad g_{ti} = \eta_{ti} + \gexp{2}_{ti} + \gexp{4}_{ti} + \dots\,, \qquad g_{ij} = \eta_{ij} + \gexp{2}_{ij} + \gexp{4}_{ij} + \dots\,. 
\end{equation}%
Here, $\eta_{\mu\nu} := {\rm diag} (-c^2,1,1,1)$ is the Minkowski background metric, and the notation $\gexp{n}_{\mu\nu}$ represents a $n$-th order quantity in the limit $1/c \to 0$, while the dots summarize higher-order terms.

Note that the first perturbation term has order $2$; except for the component $tt$, which has the dominant term of order $c^2$, and therefore its perturbation can be of order $c^0$. In fact, this perturbation is $\gexp{0}_{tt} = -2\Phi$, where $\Phi$ is the usual Newtonian potential. 

Coherently, these orders of expansion change once the $0$ index is considered instead of the $t$ one, so that:
\begin{equation}
    g_{00} = -1 + \gexp{2}_{00} + \gexp{4}_{00} + \dots\,, \qquad g_{0i} = 0 + \gexp{3}_{0i} + \gexp{5}_{0i} + \dots\,, \qquad g_{ij} = \delta_{ij} + \gexp{2}_{ij} + \gexp{4}_{ij} + \dots\,.
\end{equation}%
Newtonian dynamics is recovered in the limit $c \to \infty$ if two hypotheses are satisfied: first, if the velocity $v$ of the energy-matter content $T_{\mu\nu}$ is sub-relativistic, i.e., if $v/c \to 0$; second, if the geometry of the space-time is correctly described by the expansion (\ref{conv post newt}). The first hypothesis allows us to define the expansion (\ref{conv post newt}) in a more rigorous way, since there exists a dimensionless parameter $v/c$ going to zero, so it is meaningful to intend $\gexp{n}_{\mu\nu} \sim (v/c)^n$. When $v/c$ is small, but it is not actually zero, these higher-order corrections approximate well the general relativistic deviations from Newton theory \cite{Weinberg:1972kfs, Will_2014}.

From a mathematical point of view, it is worth asking if the off-diagonal components $g_{ti}$ are really of the second order. The expansions on the time-time and space-space components are, in fact, forced to be ``small'' compared to their non-zero dominant terms $\eta_{tt} = -c^2$ and $\eta_{ii} = 1$, so that the other terms must be of the order, respectively, smaller than $c^2$ and than $c^0$. On the other hand, there is no dominant term for off-diagonal components, since $\eta_{ti} = 0$.

From a physical point of view, one may perform a critical analysis of the hypothesis on which the Newtonian limit is based. They are intuitively expressed by saying that, in the Newtonian regime, \enquote{the speeds are small and the gravitational potentials are weak}. Although the speeds must be sub-relativistic in order to describe physical systems with low energy, it is not strictly necessary to put an additional hypothesis on the gravitational potential, i.e., on the space-time metric. On the contrary, from the first hypothesis alone one can deduce that the gravitational (and non-gravitational) forces must be ``weak'' in a suitable sense, which means that such forces cannot be strong enough to accelerate any sub-relativistic body to relativistic speed; otherwise, the first hypothesis would be broken after a time that is of the same order of the time-scale describing the given physical system.

With this in mind, it is possible to define a ``low-energy regime'' that is slightly more general than the Newtonian one. In this framework, only the gravitational forces, represented by the connection $\Gamma^{\rho}_{\mu\nu}$, are required to be small. Since they are the derivatives in space and time of the potential, given by the metric $g_{\mu\nu}$, this does not necessarily mean that the potential differences must also be small, if such differences are taken on interval of space and/or time large enough. New effects are thus allowed to emerge for big and ancient enough physical systems: this might explain why we do not observe deviations from the Newtonian paradigm for human-scale phenomena, nor for the Solar System.

A non-Newtonian low energy regime could thus be described by a non-conventional post-Newtonian expansion, such as:
\begin{equation}
\label{non conv}
    g_{tt} = -c^2 + \gexp{0}_{tt} + O(v^2/c^2), \quad g_{ti} = \gexp{0}_{ti} + O(v^2/c^2), \quad g_{ij} = \delta_{ij} + \gexp{2}_{ij} + O(v^4/c^4)\,.
\end{equation}%
As we anticipated, the time-space components are no longer restricted to be of order two. If expressed with the $0$ indices, they can be nevertheless provided with a perturbation $g_{0i} = \gexp{1}_{0i} + O(v^3/c^3)$, although the dominant term is allowed to be of the first order. If expressed with the $t$ indices, there does not seem to exist a perturbation at all, so that the dominant background geometry is no more Minkowskian. Such a non-Minkowskian background carries its own energy, momentum, and angular momentum that can affect the dynamics of matter in a non-negligible way \cite{Astesiano:2021ren, Galoppo_2024b}.

As mentioned above, these general relativistic realizations are likely to require extended energy-matter sources. Since this is the case for galaxies and galaxy clusters, but not for planetary systems and star clusters, it was proposed that this effect can be related to the problem of missing mass, which is usually ascribed to dark matter. It may also emerge from a non-negligible off-diagonal metric component or, in other words, from a general relativistic frame-dragging.

The first general-relativistic toy model for a disk galaxy was presented in the seminal work \cite{Balasin:2006cg} and it recently found a good fit from the GAIA data \cite{Crosta_2020, Beordo_2024}. The many unphysical features of this first model \cite{Ciotti:2022inn, Glampedakis_2023, Lasenby_2023, Costa_2023}, 
starting from its unrealistic rigid rotation, were overcome by more developed models. The exact GR family of solutions for a non-rigidly rotating disk galaxy was presented in \cite{Astesiano:2021ren} and then employed in \cite{re2022,Re_2024,GW2025}. The further generalization for a non-zero pressure can be found in \cite{Galoppo:2024ttc}.

A strictly related approach is the study of the gravitomagnetism of galaxies \cite{GM1, GM2, GM3, Costa_2024}. Writing a stationary metric:
\begin{equation}
\label{def GM}
    g_{tt} := -c^2 e^{2\Phi/c^2}, \quad g_{ti} := -e^{2\Phi/c^2}\xi_i\,, \quad g_{ij} := h_{ij} - c^{-2}e^{2\Phi/c^2}\xi_i\xi_j\,,
\end{equation}%
the gravitoelectric and gravitomagnetic fields can be defined as:
\begin{equation}
    \textbf{G} := -\grad\Phi\,, \quad \textbf{H} := -\frac{1}{2}e^{\Phi/c^2}\grad\times\boldsymbol{\xi}\,,
\end{equation}%
so that the geodesic equation resembles the classical electromagnetic Lorentz force:
\begin{equation}
\label{eqmot:nlGM}
    h_{ia}\partial_{\tau}U^a + \Gamma_{ijk}U^j U^k = \gamma(\gamma\textbf{G} + \textbf{U}\times\textbf{H})_i,
\end{equation}%
where $\gamma(U) := u^{\mu}U_{\mu}$ is the conserved quantity guaranteed by the normalized Killing vector $u^{\mu} = e^{-\Phi/c^2}\partial_t$. The field equations are as follows:
\begin{equation}
\label{eq:nlGM}
    \begin{cases}
        \grad\cdot\mathbf{G}= -4\pi G\rho +2\mathbf{H}^2 + \mathbf{G}^2/c^2\,, \qquad
        \grad\times\mathbf{G}=0\,,\\
        \grad\cdot\mathbf{H}=-\mathbf{G}\cdot\mathbf{H}/c^2\,, \qquad
        \grad\times\mathbf{H}=-8\pi G\mathbf{j}/c^2 + 2\mathbf{G}\times\mathbf{H}/c^2\,,
    \end{cases}\,
\end{equation}%
where $\textbf{j}$ is the density of momentum. If considered for their linear terms only, they show a form analogous to that of the Maxwell Equations.

Almost all the objections raised against the frame-dragging sustained galaxy rotation curves \cite{Ciotti:2022inn, Glampedakis_2023, Lasenby_2023, Costa_2023} are only valid for the original rigidly-rotating toy model, or for the linearized version of gravitomagnetism, and are solved in the more developed, non-linear realizations. The only criticisms that apply to any frame-dragging model appeared in \cite{Costa_2024, costa2025relativisticeffectsexplaingalactic}, but were rebutted in \cite{Re_2024}.

In these papers, the complete Einstein equations for a disk galaxy source are given. For comparison with \ref{line element} and \ref{def GM}, we write here the typical metric:
\begin{equation}
\label{eta H}
    ds^2(r,z) = -c^2 e^{2\Phi/c^2}(dt + rv_D d\phi/c^2)^2 + e^{-2\Phi/c^2}[r^2 d\phi^2 + e^{2k/c^2}(dr^2 + dz^2)]\,.
\end{equation}%
For the last, more complicated models, analytical solutions seem impossible to be found, so that the slowness of the galaxy rotation speeds, typically of $v \sim 10^{-3}c$, is exploited to simplify the calculations. In \cite{Re_2024} and \cite{Galoppo:2024ttc}, the field equations and the equations of motion are approximated to their dominant term in the limit $c\to\infty$. This is the low-energy regime that we mentioned above. Indeed, the resulting equations are not Newtonian, instead showing additional contributions due to the frame-dragging term $g_{t\phi} = -rv_D + O(v^2/c^2)$. This $v_D$ is the speed of the frame-dragging and $\text{\L}_D := rv_D$ is interpreted as the angular momentum density of the background space-time \cite{Galoppo:2024ttc}.

J. Ehlers indicated a more straightforward way to obtain the same simplifications \cite{Ehlers:2019aco}. With his Frame Theory, he unified both General Relativity and Newtonian gravity in the same coherent formalism. The axioms of the Frame Theory are expressed in terms of the parameter $\lambda = 1/c^2$, so that for $\lambda\neq0$ it is identical to GR, while the case $\lambda = 0$ means to take the low-energy limit, for which only the dominant terms of the GR equations survive.

However, in agreement with \cite{Re_2024, Galoppo:2024ttc}, for $\lambda = 0$ the Newtonian theory is not recovered, but a slightly more general one: the Newton-Cartan theory. In addition to the usual gravitational field \textbf{g}, Newton-Cartan theory displays an additional one: the Coriolis field $\boldsymbol{\omega}$. In this framework, the field equations are as follows:
\begin{equation}\label{NewtonCartanfieldeq}
\begin{cases}
    \grad\times\boldsymbol{\omega} = 0\,, &\qquad \grad\cdot\boldsymbol{\omega} = 0\,,\\
    \grad\times\mathbf{g} + 2\dot{\boldsymbol{\omega}} = 0\,, &\qquad \grad\cdot\mathbf{g} = -4\pi G\rho + 2\omega^2\,,
\end{cases}
\end{equation}%
and the equation of motion is the following:
\begin{align}
\label{NewtonCartaneom}
    \dot{\mathbf{v}} = \mathbf{g} + 2\mathbf{v}\times\boldsymbol{\omega}\,,
\end{align}%
showing a formal analogy with the gravitomagnetic equations (\ref{eqmot:nlGM}, \ref{eq:nlGM}).
The \enquote{Coriolis} field $\boldsymbol{\omega}$ is dubbed in this way because of the last equation. One can appreciate how the Field Equations boil down to Newtonian gravity when $\boldsymbol{\omega}$ vanishes, so that the last Field Equation becomes the Poisson Equation. If $\boldsymbol{\omega}$ is present, it can be considered a \enquote{modified Poisson Equation}.

\section{Exact solutions for the dark Coriolis fields}

In the past recent years, the most studied application to the real world of strong frame-dragging metrics, i.e., of non-linear gravitomagnetism, is that to the disk galaxies and to their flattish rotation curves. It was found \cite{Astesiano:2021ren, re2022} that the presence of a frame-dragging vortex can affect the distribution of matter required to sustain a given rotation curve. In \cite{Re_2024, Galoppo:2024ttc} it was explicitly shown the profile $v_D(r,0)$ that such a dragging speed needs to have on the galactic plane to substitute the role usually attributed to dark matter. We can denote ``dark frame dragging'' those particular dragging solutions with a profile suitable to have an effect analogous to dark matter.

For any observed rotation curve, being flattish or with any other profile, it is always possible to find a solution of the purely mathematical problem explaining such curve as sustained, partially or totally, by a suitable dragging vortex, if such a vortex is described only on the galactic plane. This great arbitrariness, which was successfully employed in \cite{Re_2024, Galoppo:2024ttc}, is allowed because the $z$ dependence of the frame dragging is not taken into account. Therefore, in these works, no boundary conditions for $z \to \pm\infty$ are posed, so that any choice for $v_D(r,0)$ can be used as the initial condition of the Cauchy Problem.

The physical likelihood of a frame-dragging sustained galaxy would rather descend from solving in a unique way - except for some numerical parameters - the Field Equations for $g_{ti}$ and then employ that to correct the rotation curves with respect to the Newtonian prediction. Such an exact solution is quite difficult to find for the complete, non-linear Einstein equations. One possible alternative is to follow a perturbative approach: one starts at the zeroth order by solving the Newton-Cartan field equations and then corrects the solution with higher-order terms given by a post-Newtonian-like expansion. We can dub the latter as a post-Newton-Cartan expansion.

Let us cast the Newton-Cartan equations \eqref{NewtonCartanfieldeq} and \eqref{NewtonCartaneom}, with the simplifying assumptions that the disk galaxy is essentially stationary and is symmetric with respect to its rotation axis and its galactic plane:
\begin{equation}
\label{newton cartan galaxy}
\begin{cases}
    \grad\times\boldsymbol{\omega} = 0\,, &\qquad \grad\cdot\boldsymbol{\omega} = 0\,,\\
    \grad\times\mathbf{g} = 0\,, &\qquad \grad\cdot\mathbf{g} = -4\pi G\rho + 2\omega^2
\end{cases}.
\end{equation}%
At this zeroth order, the frame dragging solution is therefore described by the Coriolis field.

In addition to the zero curl of $\textbf{g}$ implying the existence of its potential $\textbf{g} = -\grad \Phi$, the zero divergence of $\boldsymbol{\omega}$ implies the existence of a vector Coriolis potential. We recognize that such a vector potential has the dimensions of a speed. Indeed, it is $-\textbf{v}_D/2$, where we allow for now the dragging velocity field $\textbf{v}_D$ to have any direction. We can thus explicitly write the Coriolis field as follows:
\begin{equation}
\label{vector Coriolis potential}
    \boldsymbol{\omega} = -\frac{1}{2}\grad\times\textbf{v}_D = -\frac{1}{2}\left[-v_{D\phi,z}\hat{r} + (v_{Dr,z} - v_{Dz,r})\hat{\phi} + \frac{1}{r}(rv_{D\phi})_{,r}\hat{z}\right]\,,
\end{equation}
since the $\phi$ derivatives vanish due to axial symmetry. To ask that this field has no curl means:%
\begin{equation}
    0 = -2\grad\times\boldsymbol{\omega} \Rightarrow \begin{cases}
        v_{Dr,z} - v_{Dz,r} = C(r) \\
        v_{D\phi,zz} + \partial_r(v_{D\phi,r} + v_{D\phi}/r) = 0 \\
        r(v_{Dr,z} - v_{Dz,r}) = K(z)
    \end{cases} \Rightarrow \begin{cases}
        \omega_{\phi} = v_{Dr,z} - v_{Dz,r} = K/r \\
        \hat{\Delta}(rv_{D\phi}) = 0
    \end{cases}.
\end{equation}
Here, $K$ is an integration constant and $rv_{D\phi} := \text{\L}_D$ is the well-known quasilocal angular momentum (along $z$) of a dragging metric \cite{Galoppo:2024ttc}, required to satisfy the Grad-Shafranov equation, since it is the translation of the vanishing curl of $\boldsymbol{\omega}$ in terms of $\text{\L}_D$ \cite{NonZeroCoriolis}. We found thus that, in general, a stationary axisymmetric Coriolis field must have the form:%
\begin{equation}
\label{coriolis axisym}
    \boldsymbol{\omega} = -\frac{1}{2}\grad\times\left(\frac{\text{\L}_D}{r}\hat{\phi}\right) + \frac{K}{r}\hat{\phi}\,, \quad \mbox{ s.t. } \hat{\Delta}\text{\L}_D = 0\,.
\end{equation}
We want to find a solution, or a family of solutions, for this Coriolis field such that if substituted into the modified Poisson equation and into the equation of motion (including the pressure gradient):
\begin{equation}\label{eomwithpressure}
    \textbf{g} + 2\dot{\textbf{x}}\times\boldsymbol{\omega} - \frac{\grad p}{\rho} = \ddot{\textbf{x}} = -\frac{v^2}{r}\hat{r}\,,
\end{equation}%
it sustains a rotation curve $v(r)$ that is sensibly higher than those predicted by the Newtonian equations, namely:
\begin{equation}
\label{Newt eqs}
    \grad\cdot\mathbf{g}_N = -4\pi G\rho, \qquad \textbf{g}_N - \frac{\grad p}{\rho} = -\frac{v_N^2}{r}\hat{r}\,.
\end{equation}
We denote such solutions as ``Dark Coriolis Field''.

Following the very same derivation of \cite{Galoppo:2024ttc}, and expressing the quantity $\text{\L}_D$ in terms of $\boldsymbol{\omega}$ with the translation rules found in \cite{NonZeroCoriolis}, the galaxy model \eqref{eta H} is coherent with the Newton-Cartan description if
\begin{equation}
    \omega_z = -\frac{\text{\L}_{D,r}}{2r}, \quad \omega_r = \frac{\text{\L}_{D,z}}{2r}\,.
\end{equation}%
We can see that the dark Coriolis field is linked to the observed rotation curve as:
\begin{equation}
\label{dCf}
    \left[2r\omega_z^2 + 2\omega_z\partial_r(rv) + \partial_r(v^2 - v_N^2)\right]_{z=0} = 0\,.
\end{equation}%
In \cite{Galoppo:2024ttc} this identity is solved for $\omega_z$, since the scope of that work was to deduce by reverse engineering the dark Coriolis field required to sustain the rotation curve. Now we want to employ the equation in the opposite way: by substituting an exact solution for the equations (\ref{newton cartan galaxy}) and then derive the resulting rotation curve, checking if it is similar to a galactic one.

In the Newton-Cartan theory, the Coriolis field must obey linear and sourceless Field Equations: the source and the nonlinear terms will emerge as higher order contributions. This poses a mathematical dilemma with respect to the boundary conditions.

A galaxy is usually modeled as an isolated object, so it is conventional to require zero boundary conditions at spatial infinity. Such a prescription, together with the linearity and zero source of the Newton-Cartan equations, makes any regular Coriolis field vanish. To get a non-zero Coriolis field, one must either admit some singularity for the Coriolis field, or relax the boundary conditions in some way. This result for isolated objects was already highlighted in \cite{Ehlers:2019aco}, but Ehlers himself did not find it satisfactory enough to completely exclude the existence of a Coriolis field.

We want to stress here that this no-go theorem has only mathematical validity. From a physical point of view, both its hypotheses, the zero boundary conditions and the regularity of the Coriolis field, may fail. A real galaxy is not totally isolated, since it confines with other galaxies and belongs to galaxy clusters, super-clusters and bigger structures. Therefore, it is not clear if the correct boundary condition to ask is the zero one: a non-trivial Coriolis profile beyond the galactic halo may merge with the other Coriolis profiles of the surrounding galaxies.

On the other hand, a singularity in the Coriolis solutions does not necessarily mean an actual singularity, since Newton-Cartan theory is just an approximation for $\lambda = 0$. Such an approximation is reliable within a certain field of application, which provides, among other things, a fairly weak Coriolis field. If a real galaxy displays a region where the gravitomagnetism is too strong, the Newton-Cartan equations fail in describing it, since non-linear terms of the form $GH/c^2$ are no more negligible. A singular solution for the Coriolis field should then be intended as valid only far enough from the singularity; the singularity issue can find a solution with the post-Newton-Cartan expansion, e.g. by finding that it is not a real mathematical singularity, but just a maximum of the field intensity with regular profile. This may be related to the presence of super-massive black holes in the core of most galaxies.

\subsection{Cylindrical and spherical harmonics}

We can find generic solutions for the Coriolis field in the galactic case by looking at (\ref{coriolis axisym}). We ignore the singular term $K/r$ at this stage. The problem is therefore reduced to the solution of the Grad-Shafranov equation for \L$_D$.

It is an old problem now, since this is essentially the same equation faced in \cite{Balasin:2006cg}, and even before in \cite{Cooperstock:2005qw, Cooperstock:2005ba, Cooperstock:2006dt}. The $N(r,z)$ field in the Balasin-Grumiller model coincides indeed with the \L$_D$ quantity, for that particular case. It is also the same quantity as the $\mathcal{F}$ that generates the Velocity Field Equation in \cite{Astesiano:2021ren, Re_2024}. However, it is worth noting that the Grad-Shafranov equations for $N$ or $\mathcal{F}$ are exact equations, because of the simplifying, unphysical assumptions of rigid rotation and/or zero pressure employed in those models; while the Grad-Shafranov equation for \L$_D$ is just the zeroth-order approximation, destined to be surpassed in the post-Newton-Cartan expansion.

Following \cite{Balasin:2006cg, NonZeroCoriolis}, the harmonic solutions in cylindrical coordinates are described by Bessel functions:
\begin{equation}
    \text{\L}_D(r,z) = \lambda r[A(\lambda)K_1(\lambda r) + B(\lambda)I_1(\lambda r)]\cos(\lambda z)\,,
\end{equation}%
and the generic solution can be built up by integrating them in $d\lambda$. The solutions in $\sin(\lambda z)$ are excluded by the requirement of planar symmetry (symmetry under the transformation $z \to -z$).

This form of the solution is very useful when evaluated on the galactic plane $z=0$, so that one chooses the coefficients that fit the observed rotation curve through (\ref{dCf}). This is the strategy, we mentioned above, of taking \L$_D(r,0)$ as the initial value and generating from it the Coriolis field on the whole space, ignoring the boundary behavior and any singularity.

Since we want here to study the behavior at spatial infinity, no matter if along $r$ or $z$, it may be more useful to express the equations in spherical coordinates $(R, \theta, \phi)$, where $R^2 = r^2 + z^2$ and $\sin\theta = z/R$. The dependence $\phi$ vanishes as well, because of axial symmetry. We can thus try the separation of variables \L$_D(R,\theta) := f(R)g(\sin\theta)$, so that we find:
\begin{align}
    0 = \hat{\Delta}\text{\L}_D = \partial_R^2\text{\L}_D + \frac{1}{R^2}\partial_{\theta}^2\text{\L}_D + \frac{\tan\theta}{R^2}\partial_{\theta}\text{\L}_D \Rightarrow R^2\frac{f''(R)}{f(R)} = -\cos^2\theta\frac{g''(\sin\theta)}{g(\sin\theta)} = \alpha(\alpha-1)\,,
\end{align}
whose solution is:
\begin{align}
\label{sph harm}
     &\text{\L}_D(R,\theta) = \left[A_+(\alpha)R^{\alpha} + A_-(\alpha)R^{1 - \alpha}\right]\times\\
    &\times\left[B_0(\alpha)\,_2F_1(-\alpha/2, \alpha/2 - 1/2; 1/2; \sin^2\theta) + B_1(\alpha)\sin\theta\,_2F_1(\alpha/2, 1/2 - \alpha/2; 3/2; \sin^2\theta)\right].\nonumber
\end{align}
This family of solutions with hypergeometric functions is symmetric with respect to the special value $\alpha = 1/2$, for which the degenerate form $f(R) = A_0\sqrt{R} + A_1\sqrt{R}\ln R$ is found. A general solution can be found by integrating in $d\alpha$ the spherical harmonics above.

However, these harmonics nicely allow us to study the boundary behavior $R\to\infty$ as power laws, with positive or negative exponents; as well as it is easy to study any singularity in the central point $R\to0$ or on the symmetry axis $\sin^2 \theta \to 1$. On the galactic plane $\theta = 0$ the hypergeometric factor is trivial, so \L$_D(r,0) = B_0(A_+r^{\alpha} + A_-r^{1-\alpha})$ is also very easy to study. To find a dark Coriolis field within these harmonics, its component that must fit equation (\ref{dCf}) is:
\begin{equation}
\label{omega harm}
    \omega_z(r,0) = -\frac{\text{\L}_{D,r}}{2r} = -\frac{B_0}{2}[\alpha A_+r^{\alpha-2} + (1-\alpha)A_-r^{-\alpha-1}]\,.
\end{equation}%
The zero boundary condition can be met for any value of $A_-$ and $A_+$ with $-1<\alpha<2$, for $A_+=0$ with $\alpha<-1$ and for $A_-=0$ with $\alpha>2$. The point-like singularity in $r\to 0$ is avoided under the opposite conditions. Concerning the hypergeometric functions, they are regular in $\sin^2\theta\to1$ if and only if one of their arguments $\pm\alpha/2, \pm(\alpha-1)/2$ is a non-positive integer, otherwise their first derivatives display a logarithmic divergence. Therefore, the axial singularity is avoided for $B_1=0$ with $\alpha = 2n$ or $\alpha = -2n+1$, where $n\geq0$ is a natural number, and for $B_0=0$ with $\alpha = 2n-1$ or $\alpha = -2n$. 

\subsection{Features of the dark Coriolis fields in three regions}

\subsubsection{The halo}

In the halo region, we can try to explain a flat rotation curve $v(r)\equiv v_f \simeq 220$ km/s with a suitable dark Coriolis field. It must therefore fit (\ref{dCf}), which can be solved as a simple algebraic equation:%
\begin{equation}
\label{effective omega}
    -2r\omega_z(r,0) = \text{\L}_{D,r}(r,0) = (rv)_{,r} - \sqrt{(rv)_{,r}^2 - 2r(v^2 - v_N^2)_{,r}}\,,
\end{equation}%
where the solution is chosen with the minus sign so that the purely Newtonian case is recovered for $v=v_N \Leftrightarrow \omega_z=0$.

We try to assume that the only matter present is the baryonic one $\rho := \rho_B$. The halo is the region where the effect of the baryonic component should act as an approximately point-like source, so that $v_N(r)^2 \simeq GM_B/r$, with $M_B\simeq 9\cdot10^{10} M_{\odot}$ the total baryonic content, and $v_f$ is the asymptotic velocity assumed in that region. The above equation thus requires:
\begin{equation}
    \omega_z(r,0) \simeq -\frac{1}{2r}\left(v_f - \sqrt{v_f^2 - 2\frac{GM_B}{r}}\right)\,.
\end{equation}
If the flattish curve goes on until spatial infinity (as is claimed by Theories of Modified Gravity \cite{Mistele_2024}), this solution would have an asymptotic behavior:
\begin{align}
    &\omega_z \simeq -\frac{GM_B}{2v_f}r^{-2} - \frac{G^2M_B^2}{4v_f^3}r^{-3} + O(r^{-4})\,, \cr
    &\text{\L}_D \simeq \frac{GM_B}{v_f}\ln\frac{r}{r_0} - \frac{G^2M_B^2}{2v_f^3}r^{-1} + O(r^{-2})\,,
\end{align}%
where the integration constant is expressed as the arbitrary parameter $r_0$. The dominant term can be obtained as a limit of spherical harmonics (\ref{sph harm}):
\begin{equation}
    \text{\L}_D(r,0) = \lim_{\alpha\to0}\frac{GM_B}{\alpha v_f}(r^{\alpha} - r_0^{\alpha}), \quad \omega_z(r,0) = \lim_{\alpha\to0}\left(-\frac{GM_B}{2v_f}r^{\alpha-2}\right)\,.
\end{equation}%
The hypergeometric functions in \eqref{sph harm} collapse to $1$ when their first argument $\pm\alpha/2$ tends to zero, so it is sufficient to choose $B_{0,1}$ independently of the parameter $\alpha\to0$ to have a solution regular on the axis:
\begin{equation}
    \text{\L}_D(R,\theta) = \lim_{\alpha\to0}\frac{GM_B}{\alpha v_f}(R^{\alpha} - r_0^{\alpha})(B_0 + B_1\sin\theta) = \frac{GM_B}{2v_f}\left(\ln\frac{r^2+z^2}{r_0^2}\right)\left(1 + B_1\frac{z}{\sqrt{r^2 + z^2}}\right)\,.
\end{equation}%
The coefficient $B_0$ must be set to $1$ in order to recover the desired profile on the galactic plane, while the coefficient $B_1$ describes the behavior on the axis:
\begin{equation}
    \text{\L}_D(0,z) = \frac{GM_B}{v_f}\left(\ln\frac{|z|}{r_0}\right)\left[1 + B_1 Sg(z)\right]\,.
\end{equation}

\subsubsection{Beyond the halo}

A different possibility is given if the rotation curve does not maintain its flattish behavior forever, but it eventually decays Newtonianly $v^2 \simeq GM_{\rm tot}/r$, in a beyond-halo region (say, $r>50$ or $100$ kpc). If the total mass has a dark matter component, $M_{\rm tot} = M_B + M_{\rm dm}$, we can nevertheless have a dark Coriolis field that mimics it.

If this profile is substituted in (\ref{dCf}), the harmonic ansatz $\omega_z(r,0) = -\frac{1}{2}A\alpha r^{\alpha-2}$ fits it for the exponent $\alpha=1/2$ and coefficient $A = \sqrt{GM} - \sqrt{GM + 8GM_{\rm dm}}$. In other words, an exact solution is given by:
\begin{align}
    \omega_z(r,0) &= \frac{\sqrt{GM + 8GM_{\rm dm}} - \sqrt{GM}}{8}r^{-3/2}\,, \\
    \text{\L}_D(r,z) &= (\sqrt{GM} - \sqrt{GM + 8GM_{\rm dm}})(r^2 + z^2)^{1/4} \,\times\nonumber\\
    \times&\left[\,_2F_1\left(-\frac{1}{4}, -\frac{1}{4}; \frac{1}{2}; \frac{z^2}{r^2 + z^2}\right) + B_1\frac{z}{\sqrt{r^2 + z^2}}\,_2F_1\left(\frac{1}{4}, \frac{1}{4}; \frac{3}{2}; \frac{z^2}{r^2 + z^2}\right)\right]\,.
\end{align}%
It is singular on the axis, with the exact behavior of the singularity given by the parameter $B_1$.

\subsubsection{The bulge}

The third regime to be considered is that of the central bulge. The observed rotation curve in that region shows approximately a linear growth:
\begin{equation}
    v(r) = \chi_{\rm obs}r + o(r)\,.
\end{equation}%
Since the density shows a core with central peak $\rho_{B0}$, a linear growth is coherent with the Newtonian prediction (\ref{Newt eqs}), so that:
\begin{equation}
    v_N(r) = \chi_N r + o(r), \quad \chi_N^2 = \frac{4}{3}\pi G\rho_{B0}\,.
\end{equation}%
Here the pressure can be neglected since it is almost constant in the center. However, some missing mass effects are detected even in the bulge. The deduced dark matter distribution has a core, according to empirical data, so that its central density $\rho_{\rm dm0}$ is subdominant with respect $\rho_{B0}$ in most of the galaxies. However, some rarefied dwarf galaxies have $\rho_{\rm dm0} > \rho_{B0}$. The Newtonian description of these facts includes the dark matter component, so that:
\begin{equation}
    \chi_{\rm obs}^2 = \frac{4}{3}\pi G(\rho_{B0} + \rho_{\rm dm0})\,.
\end{equation}
Substituting these profiles into (\ref{effective omega}), we see that the same missing mass effects can be explained by a suitable Coriolis field, so that approaching the center it takes the definite value:
\begin{equation}
    \omega_z(0,0) = -\frac{2\chi_{\rm obs}r - \sqrt{4\chi_{\rm obs}^2 r^2 - 4(\chi_{\rm obs}^2 - \chi_N^2)r^2}}{2r} = \chi_N - \chi_{\rm obs}\,.
\end{equation}%
This non-singular behavior can be described by the harmonic solution (\ref{omega harm}) for $\alpha = 2, A_-=0$ and $A_+ B_0 = \chi_{\rm obs} - \chi_N$; or for $\alpha = -1, A_+=0$ and $A_- B_0 = \chi_{\rm obs} - \chi_N$. For both choices, singularities are avoided on the axis when $B_1=0$, so that:
\begin{equation}
    \text{\L}_D(R,\theta) = (\chi_{\rm obs} - \chi_N)R^2 \,_2F_1(-1, 1/2; 1/2; \sin^2\theta) = (\chi_{\rm obs} - \chi_N)r^2\,.
\end{equation}
If the bulge region is described by the choice $\alpha=-1$, the three regimes appear to take progressively growing values, proceeding from the center towards the outside of the galaxy: $\alpha=-1$ in the bulge region, $\alpha=0$ in the halo region, and $\alpha=1/2$ in the beyond-halo region. We can expect a transition from one to another of these three harmonic regimes to take place on the border between their appropriate regions. However, linear theory cannot describe exactly how their merging occurs. Since a non-linear theory is required, we can recover it from the higher-order terms of a post-Newton-Cartan expansion.

\section{Non-conventional post-Newtonian expansion as Newton-Cartan gravity}

It is interesting to note that equations (\ref{newton cartan galaxy}) and the equation of motion can be derived from a quite different approach based on a non-conventional post-Newtonian expansion. In the line element (\ref{line element}) we separate time and space components. Then, the geodesic equations take the following general form:
\begin{equation}\label{full geodesic eq}
    \frac{d^2x^i}{dt^2}=-\Gamma^{i}_{tt}-2\Gamma^{i}_{tj}v^{j}-\Gamma^{i}_{jk}v^j v^k+\Gamma^{t}_{tt}v^i+2\Gamma^{t}_{tj}v^iv^j+\Gamma^{t}_{jk}v^jv^kv^i\,.
\end{equation}
Our approximation scheme is based on an expansion in powers of $1/c$, according to which the components of the metric take the form given in \eqref{non conv}. We can now recognize the Newtonian potential and the dragging speed in $\gexp{0}_{tt} = -2\Phi$ and $\gexp{0}_{ti} = -v_{Di}$.

The contravariant components of the metric, expanded analogously, are found exploiting the identity relation $g^{\mu\rho}g_{\rho\nu} = \delta^{\mu}_{\nu}$. Applied to the orders $c^2$ and $c^0$, it allows us to find that:
\begin{equation}\label{PNCexpansionnexttoleading}
    g^{tt} = -c^{-2} + \gexp{4}^{tt} + O(1/c^6)\;, \quad g^{ti} = -v_D^i/c^2 + \gexp{4}^{ti} + O(1/c^6)\,, \quad g^{ij} = \delta^{ij} + \gexp{2}^{ij} + O(1/c^4)\;.
\end{equation}
The lowest-order non-vanishing Christoffel symbols that enter in \ref{full geodesic eq}, are given in terms of the components of the metric by:
\begin{equation}\label{Christoffel0}
    \OrdExp{0}{\Gamma}^{i}_{tt}=-\frac{1}{2}\frac{\partial \gexp{0}_{tt}}{\partial x^{i}} + \pdv{\gexp{0}_{ti}}{t} = \partial_i\Phi - \dot{v}_{Di} \;, \qquad
    \OrdExp{0}{\Gamma}^{i}_{tj} = \frac{1}{2}\lp \frac{\partial \gexp{0}_{it}}{\partial x^j}-\frac{\partial \gexp{0}_{jt}}{\partial x^i}\rp = \frac{1}{2}\epsilon_{ijk}(\grad\times\textbf{v}_D)^k \;.
\end{equation}
With definitions analogous to 
those in \cite{Ehlers:2019aco, NonZeroCoriolis}:
\begin{equation}\label{definitions Gamma}
    \OrdExp{0}{\Gamma}^i_{tt} := -g^i\;, \qquad \OrdExp{0}{\Gamma}^i_{tj} := -\epsilon_{ijk}\,\omega^k\;,
\end{equation}
and the geodesic equation takes at the zeroth order the same form of (\ref{eomwithpressure}) without pressure:
\begin{equation} \ddot{\mathbf{x}} = \mathbf{g} + 2\dot{\mathbf{x}}\times\boldsymbol{\omega}\;.
\end{equation}
Comparing (\ref{Christoffel0}) and (\ref{definitions Gamma}), we immediately find two vector relations that cannot be deduced in the standard Newton-Cartan theory but enrich it:
\begin{equation}
\label{beautiful relations}
    \textbf{g} = -\grad\Phi + \dot{\textbf{v}}_D, \qquad \boldsymbol{\omega} = -\frac{1}{2}\grad\times\textbf{v}_D\,.
\end{equation}%
The second one reveals that the vector potential of the Coriolis field, which we knew existed from the Newton-Cartan equations, has a physical meaning in terms of the dragging speed. Moreover, the first of \eqref{beautiful relations} shows how the Newtonian field $\textbf{g}$ is generalized in the non-stationary case, when it is no longer the pure gradient of a scalar potential.

In order to determine the field equations at the lowest order, we need first to specify the matter content. We assume it to be given by a perfect fluid, namely: $T^{\mu\nu} = (\rho + p/c^2) u^{\mu}u^{\nu} + p g^{\mu\nu}$, with $\rho$ its density, and $p$ its pressure. Then, we expand the full Einstein equations at the lowest order in $1/c$, obtaining:
\begin{equation}
    \grad\cdot \mathbf{g} = -4\pi G\rho + 2{\omega}^2, \, \qquad \grad\times\boldsymbol{\omega} = 0\;.
\end{equation}
These are exactly the second equation in the first row and the first equation in the second row of (\ref{NewtonCartanfieldeq}). Retrieving the remaining equations of (\ref{NewtonCartanfieldeq}) is immediate from \eqref{beautiful relations}.

Note that, at the dominant order of the expansion, pressure does not play a relevant role since it already enters the Einstein equations multiplied by a factor $1/c^2$. The same is also true for the components $T_{0i}$, which provide the source of the field equation containing $\grad\times\boldsymbol{\omega}$.

We note that with the non-conventional Post-Newtonian expansion that we have described we retrieve at the lowest order in $1/c$ a dynamics that is richer than the Newtonian one and which agrees with the results found in \cite{NonZeroCoriolis}. Furthermore, we expect the same expansion to reduce, in the stationary case and at the next-to-leading order, to the non-linear gravitomagnetism equations \eqref{eqmot:nlGM} and \eqref{eq:nlGM}, with the Coriolis field $\boldsymbol{\omega}$ being the zero order approximation of the gravitomagnetic field $\mathbf{H}$.

The next-to-leading order of the expansion given in \eqref{PNCexpansionnexttoleading}, and its applications, will be investigated in more detail in future papers.

\section{Discussion and Conclusions}

In this paper, we have investigated an application of Newton-Cartan gravitational theory to the velocity curves of disk galaxies. In Newton-Cartan theory, a new field, in addition to the usual one, is present: the Coriolis field. In the Introduction, we have presented the non-linear version of the gravitomagnetic theory and have shown its compatibility with the Newton-Cartan theory since they are simplified versions of General Relativity under different assumptions: under stationarity and under the limit of $1/c$ going to zero, respectively.

In the second section, we have explicitly solved the field equation for the Coriolis field, essentially a Laplace equation, in spherical coordinates. Then, we have applied the solution for three regimes of interest in describing the typical dynamics of a disk galaxy, namely the bulge region, the flattish velocity profile region, and the faraway region, where we expect the velocity field to decay. The solution found adapts nicely to the three regimes, providing support for the presence of a Coriolis field. Here, we have dubbed \enquote{Dark Coriolis field} such profiles, since they are able to mimic Dark Matter phenomena. The issue of gluing together the three regimes still remains and will be addressed in future investigations. 

Finally, we have shown how to frame the Newton-Cartan gravitational theory in a more general one via an expansion similar to the standard post-Newtonian, but with the presence of a non-vanishing space-time component of the metric, which, at the leading order, is of the same order as the usual gravitational potential. The ensuing expansion was dubbed the \enquote{Non-Conventional} Post-Newtonian expansion and will be developed and studied more extensively in its implications in future investigations.

\bibliographystyle{unsrt}
\bibliography{NCPNEx.bib}

\end{document}